\documentclass[twocolumn,preprintnumbers,amsmath,amssymb]{revtex4-2}

\usepackage{graphicx}
\usepackage{dcolumn}
\usepackage{bm}

\usepackage{lineno}

\newcommand{\abs}[1]{|#1|}

\newcommand{\scal}[2]{\langle#1|#2\rangle}

\usepackage{verbatim}

\begin{document}

\title{One-dimensional Dexter-type excitonic topological phase transition}

\author{Jianhua Zhu$^{1,2}$}\email{ucapjhz@ucl.ac.uk}
\author{Ji Chen$^{1,3,4}$}\email{ji.chen@pku.edu.cn}
\author{Wei Wu$^2$}\email{wei.wu@ucl.ac.uk}
\affiliation{$^1$School of Physics, Peking University, Chengfu Road 209, Haidian, Beijing 100871, China}

\affiliation{$^2$UCL Department of Physics and Astronomy, University College London, Gower Street, London WC1E 6BT}
\affiliation{$^3$Interdisciplinary Institute of Light-Element Quantum Materials and Research Center for Light-Element Advanced Materials, Peking University, Beijing 100871, P. R. China}
\affiliation{$^4$Frontiers Science Center for Nano-Optoelectronics, Peking University, Beijing 100871, P. R. China}

\date{\today}

\begin{abstract}
Recently topogical excitons have attracted much attention. However, studies on the topological properties of excitons in one dimension are still rare. Here we have computed the Zak phase for a generic one-dimensional dimerised excitonic model. Tuning relevant hopping parameters gives rise to a rich spectrum of physics, including non-trivial topological phase in uniform chain unlike the conventional Su-Shcrieffer-Heeger model, topologically nontrivial flat bands, and exotic fractional phase. a new concept of ``composite chiral site" was developed to interpret the Zak phase of $\pi$ in our calculations. Our finite-chain calculations substantiate topological edge states, providing more information about their characteristics. Most importantly, in the first time, a topological phase transition assisted by the Dexter electron exchange process has been found. 
\end{abstract}

\maketitle

\section{Introduction}
\begin{figure}[htbp]
\includegraphics[width=6.5cm, height=7.cm, trim={0cm 0cm 0.0cm 0.0cm},clip]{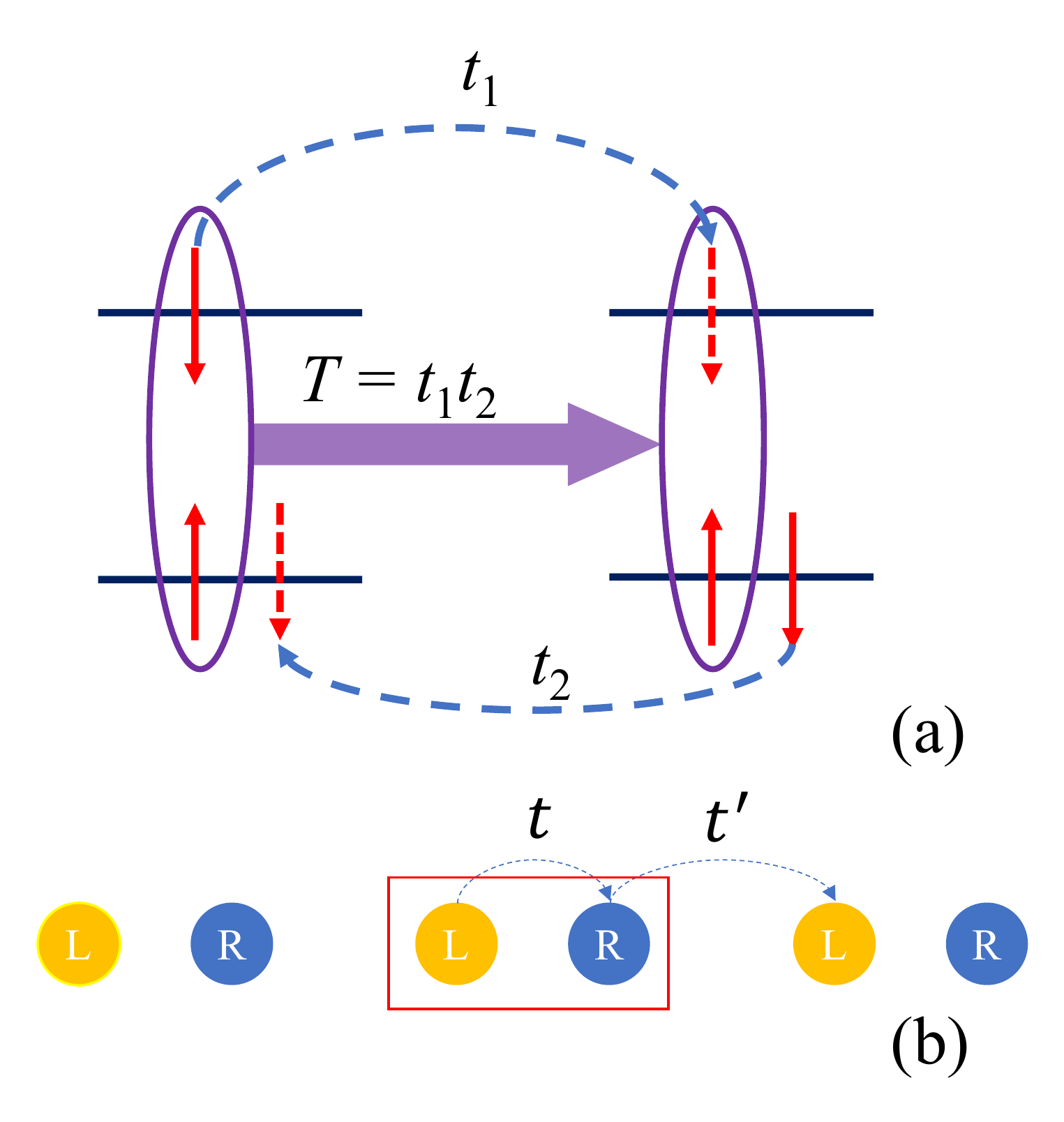}\\
\caption{(a) The Dexter electron exchange is illustrated. $T=t_1t_2$ is the effective hopping amplitude for the entire single exciton. (b) The SSH model consists of L and R sub-lattices, intra-cell hopping ($t$), and inter-cell hopping ($t^\prime$).}\label{fig:dexter}
\end{figure}

Exciton - a bound state between electron and hole - plays a vital role in optics, which can be either delocalised (Wannier-Mott exciton) or localized (Frenkel exciton) \cite{kittel2018}. The Frenkel exciton, normally formed in insulator or molecular crystals, migrates through two hopping mechanisms, namely F\"oster coupling and Dexter electron exchange \cite{deng2010exciton,anantharaman2021exciton,dresselhaus2007exciton,menke2014exciton,mikhnenko2015exciton,lee2013singlet,xu2021recent}. Dexter electron exchange can be seen as simultaneous hopping of electrons on the upper and lower levels, resulting in a migration of entire exciton (Fig.\ref{fig:dexter}a). Excitonic migration from the 'antenna' pigment to the reaction centre is of great importance in plant photosynthesis that fuels the livings \cite{zhang2019artificial,stirbet2020photosynthesis,sarovar2010quantum}. The recent topical research - topological photonics (TP) attempts to construct topological states by engineering light \cite{ozawa2019topological,lu2014topological,ota2020active,kim2020recent,segev2020topological,wu2017applications,price2022roadmap}. On the other hand, topological edge state could also arise from topology of interactions based on natural setup without any artificial engineering. Locally excited (LE) and charge-transfer (CT) states (formed by the electron and hole on different sites) could occur in a wide range of physical systems. Especially CT process is crucial both for lives on Earth and our modern daily life relying on electricity \cite{derr2020multifaceted}. Recent research suggests that CT state could play an important role in organic light emitting diode, solar cells, and resistive memory devices \cite{cui2020fast,han2019local,yam2020charge}. CT excited states can also mediate the transition in the Dexter electron exchange process \cite{osti_6089580}. 

One-dimensional (1D) chain structures, which could be formed by atoms, molecules, quantum dots, or semiconductor dopants, have recently attracted much attention due to their interesting topological properties \cite{crain2005end,hirjibehedin2006spin,khajetoorians2019creating,choi2019colloquium,zhao2023quantum,r2020conducting,nichol2022quantum,kiczynski2022}. The Su-Shcrieffer-Heeger (SSH) model is a classic example for topological phase transition in 1D (Fig.\ref{fig:dexter}a) \cite{su1979solitons,su1980soliton}. The SSH model consists of two sub-lattices (L and R) in Fig.\ref{fig:dexter}(b), which respects chiral symmetry, leading to integer winding number ($\mathbb{Z}$-invariant). The corresponding Zak phase (the Berry's phase for 1D) can therefore take integer number of $\pi$. The inversion symmetry further constrains the Zak phase to be either odd or even multiples of $\pi$ \cite{zak1989berry,asboth2016schrieffer}, rendering a $\mathbb{Z}_2$ invariance. In the SSH model, when the hopping strength between cells is greater than that within the cell, there appears a topologically non-trivial phase, with a Zak phase of $\pi$. The SSH model was first proposed in polyacetylene polymer chain with alternating single- and double-carbon bonds, but could also be realized in the dopant chains in semiconductors \cite{zhu2020linear,zhu2021multihole,wu2020optical}. The SSH model has been realized in TP as reviewed in details in Ref.\cite{ota2020active}. A fascinating topic for TP in 1D is flat band, which has recently stimulated intense research interest in condensed matter physics, photonics, and meta-materials \cite{leykam2018, bergholtz2013topological,balents2020superconductivity,vicencio2021photonic,poblete2021photonic,derzhko2015strongly,liu2014exotic,tang2020photonic}. Flat bands also have many potential applications such as slow light due to the vanishing group velocity \cite{leykam2018}. In addition, one-dimensional systems could be useful for the demonstration of topological order, charaterized by a phase transition beyond Landau's symmetry-breaking principle \cite{wenbook}. To the authors' best knowledge, studies on topological properties of 1D excitons are still rare although topological excitons have been discussed extensively in two dimensions \cite{wu2017topological,gao2018controlled,varsano2020monolayer,wang2023excitonic,mori2023spin}.


Previously we have established a generic excitonic model for one-dimensional molecular chains, which took into account both the intramolecular LE and inter-molecular CT excited states that could interpret a body of optical experimental results \cite{chen2021optoelectronic}. Both of these excited states are important for the optical and charge dynamics in optics \cite{armin2021history,wang2021recent}. More importantly, this model can be generalized to any one-dimensional physical system formed by atoms, molecules, quantum dots, impurities, and so on \cite{kiczynski2022, le2020topological,zhu2020linear,zhu2021multihole,wu2020optical}. For the details of the model Hamiltonian please see Ref.\cite{chen2021optoelectronic}. It would therefore be necessary to understand this model systematically. Here we have mapped the six excitonic states, LE1, LE2, CT1, CT2, CT3, and CT4, in the unit cell to better illustrate the couplings between them, as shown in Fig.\ref{fig:2}(a), where we have also labeled the states by A-F for the convenience of discussion. The Hamiltonian reads
\begin{eqnarray}
\hat{H} =\sum_n &&t_2 a^{\dagger}_{\mathrm{LE1},n}a_{\mathrm{CT1},n} + t_1 a^{\dagger}_{\mathrm{LE1},n}a_{\mathrm{CT2},n}+\nonumber\\
&& t_1^\prime a^{\dagger}_{\mathrm{LE1},n}a_{\mathrm{CT3},n-1} + t_2^{\prime} a^{\dagger}_{\mathrm{LE1},n}a_{\mathrm{CT4},n-1}+\nonumber\\
&&t_1 a^{\dagger}_{\mathrm{LE2},n}a_{\mathrm{CT1},n} + t_2 a^{\dagger}_{\mathrm{LE2},n}a_{\mathrm{CT2},n}+\nonumber\\
&& t_2^\prime a^{\dagger}_{\mathrm{LE2},n}a_{\mathrm{CT3},n} + t_1^\prime a^{\dagger}_{\mathrm{LE2},n}a_{\mathrm{CT4},n}+\nonumber\\
&&-\frac{d}{4}a^{\dagger}_{\mathrm{LE1},n}a_{\mathrm{LE1},n}-\frac{d}{4}a^{\dagger}_{\mathrm{LE2},n}a_{\mathrm{LE2},n}+\nonumber\\
&&\frac{d}{4}a^{\dagger}_{\mathrm{CT1},n}a_{\mathrm{CT1},n}+\frac{d}{4}a^{\dagger}_{\mathrm{CT2},n}a_{\mathrm{CT2},n}+\nonumber\\
&&\frac{d}{4}a^{\dagger}_{\mathrm{CT3},n}a_{\mathrm{CT3},n}+\frac{d}{4}a^{\dagger}_{\mathrm{CT4},n}a_{\mathrm{CT4},n}+\nonumber\\
&& +\  h. c.
\end{eqnarray}\label{eq:cth0}

This Hamiltonian can be transformed to the momentum space, as follows.
\begin{equation}\label{eq:cth}
\hat{H}_k=\begin{pmatrix}
-\frac{d}{2}&0&t_2&t_1&t_1^\prime e^{-ik}&t_2^\prime e^{-ik} \\
0&-\frac{d}{2}&t_1&t_2&t_2^\prime&t_1^\prime \\
t_2&t_1&\frac{d}{2}&0&0&0 \\
t_1&t_2&0&\frac{d}{2}&0&0\\
t_1^\prime e^{ik}&t_2^\prime&0&0&\frac{d}{2}&0 \\
t_2^\prime e^{ik}&t_1^\prime&0&0&0&\frac{d}{2}
\end{pmatrix}
\end{equation}

Due to the nature of optical excitation, which has to connect two states, the unit cell of the model is asymmetric, i.e., generally breaking inversion symmetry. The couplings $t_1$, $t_2$ have been differentiated from $t_1^\prime$ and $t_2^\prime$ to account for dimerisation. We set $d=0$ in most of the paper as it doesn't affect very much the qualitative picture. We adopt the sequence $\{t_1,t_2,t_1^\prime,t_2^\prime\}$ to describe our parameter set. Our calculation results are not only consistent with the previous experimental and numerical works \cite{reid2022}, but also, more importantly can lead to a rich spectrum of physics, ranging from fractional phase to topologically nontrivial flat bands. 

\section{Results and Discussions}\label{sec:results}

\subsection{Topological phase induced by decoupling and the composite chiral site}

The results for only turning on one or two parameters are shown in the supplementary information (SI). For one-parameter model, we can obtain the topologically non-trivial state with similar mathematical structure as compared with the SSH model. For two-parameter model, we have seen the interference between the topological states, and the Zak phase of the flat bands is qualitatively consistent with the previous work \cite{reid2022}. Here we compute the total Berry phase for the degenerate multi-band case \cite{vanderbilt2018}.  

\begin{figure*}[htbp]
\includegraphics[width=17cm, height=9cm, trim={0cm 0cm 0.0cm 0.0cm},clip]{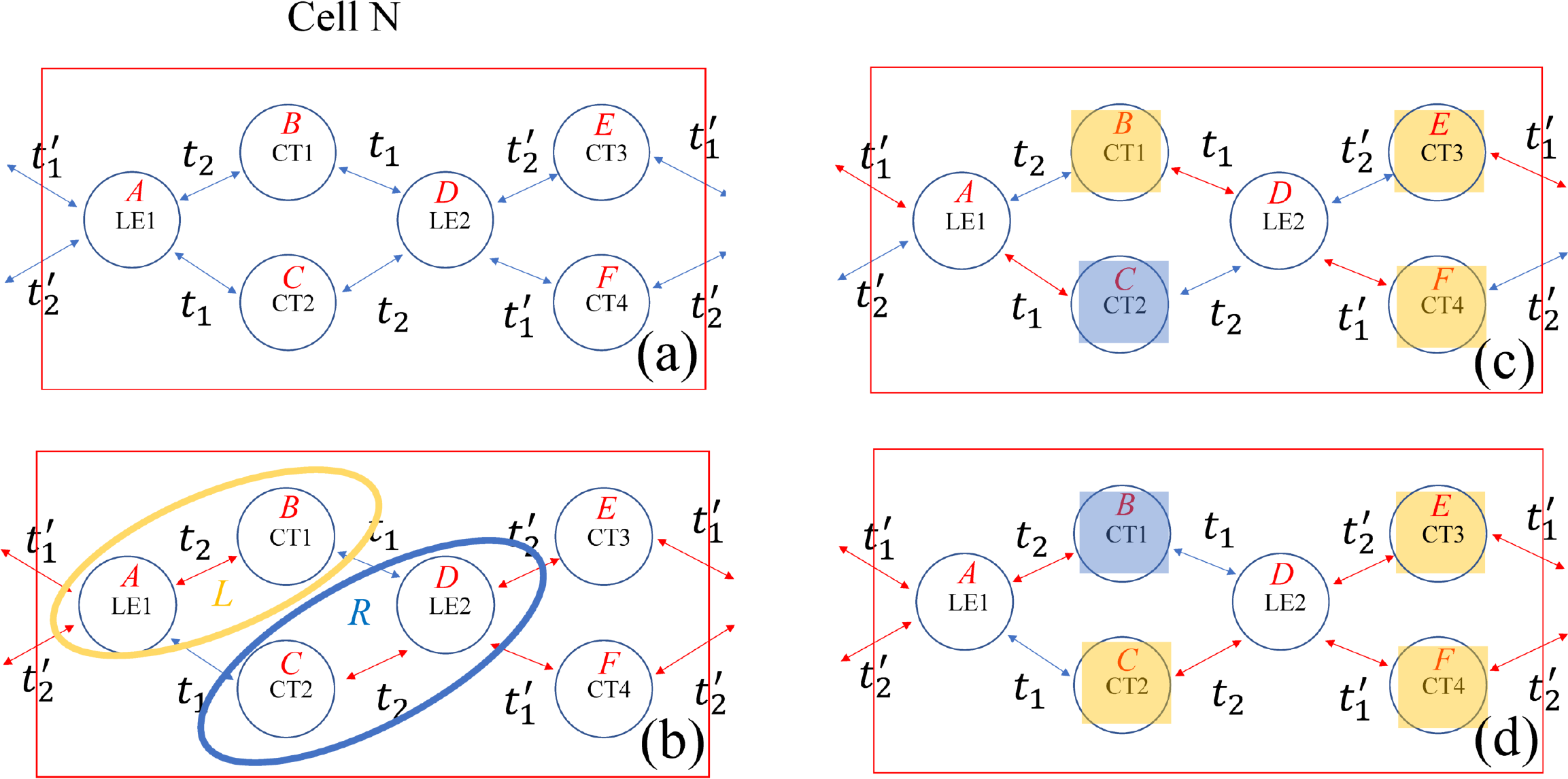}\\
\caption{(a) We have mapped the model diagram in Ref.\cite{chen2021optoelectronic} to a state-based structure with couplings and state labels. (b) An illustration of the two composite chiral sites for $t_1=0,t_2=1,t_1^\prime=t_2^\prime=x\neq 0$: the left-hand state ($L$) formed by LE1 and CT1 coupled by $t_2$ (circled in yellow) and the right-hand state ($R$) formed by LE2 and CT2 (circled in blue). $L$ and $R$ are decoupled by $t_1$. We have also shown an illustration of the components in the eigenvectors for for $\{ 1,0,1,0\}$ (c) and $\{0,1,x,y\}$ with $\abs{x^2-y^2} = 1$ (d). Here we use red (blue) colour to represent non-zero (zero) hopping. $L$ state is in yellow, while $R$ state is in blue.}\label{fig:2}
\end{figure*}

We have computed all the twelve combinations for the four hopping parameters in the set of $0, 1, x$, and $y$, in which we have provided the analytical formulae for all the Zak phases in the SI. For the three-parameter model all the bands are flat owing to decoupling. We show here an interesting example for $\{0, 1, x, y\}$ in Table.\ref{tab:3}.  The flat bands at energy zero could be topologically nontrivial when $\abs{x^2-y^2}=1$. Especially when $x=y$, the Zak phases of the third and fourth bands is $\pi$; the two associated eigenvectors are $v_3=\frac{1}{2}[e^{-ik},-1,-e^{-ik},1,0,0]$ (the anti-bonding state between LE and CT) and $v_4=\frac{1}{2}[-e^{-ik},1,-e^{-ik},1,0,0]$ (the bonding state of LE and CT). Based on the mathematical structure of the eigenvectors, the group of the sites occupied by the linear combination of excitonic states can be defined as a ``composite chiral site" (CCS), as shown in Fig.\ref{fig:2}(b), thus leading to a phase of $\pi$. This concept is proposed in the first time for the case where there are more than two sites in a unit cell. We can see that there would be a minor sign if exchanging A, D and B, C for $v_2$ simultaneously at the $\Gamma-$point. The left-hand CCS ($L$) is therefore formed by LE1 and CT1 coupled by $t_2$ and $t_2^\prime$ while the right-hand CCS ($R$) is formed by LE2 and CT2. These two CCSs are decoupled by $t_1$, which is the key to resume the chiral symmetry and render the topologically non-trivial band. Similarly when $x = -y$, the fifth and sixth bands are topologically nontrivial. In Fig.2 of SI, we have shown more examples for the scenarios with the Zak phase of $\pi$, which share the similar features as discussed above. 

Regarding the Zak phase for the zero-mode flat bands, we have analysed the associated eigenvectors for $\{ 1,0,1,0\}$ and $\{0,1,x,y\}$ with $\abs{x^2-y^2} = 1$; in both scenarios, the sum of the phases for the flat bands is $\pi$ (Fig.\ref{fig:2}c). These flat bands are formed by four CT states. We can also implement the concept of CCS coupled by non-zero hopping, which are decoupled by other parameter to form L and R states. As shown in Fig.\ref{fig:2} (d), $L$ CCS is formed by CT2-4 (coupled by $t_1^\prime$, $t_2$, and $t_2^\prime$), and $R$ is formed by CT1, both of which are decoupled by $t_1$. In summary, we can see that (i) when we can clearly decouple group of states within a cell, we can have the phase of $\pi$ assisted by the symmetry of hopping parameters and (ii) once breaking the symmetry of hopping parameters, fractional phase will appear. 


\begin{table*}[htbp]
\caption{Our calculated Zak phases for $t_1 = 0$,  $t_2 = 1$, $t_1^\prime = x$, and $t_2^\prime = y$.}
\begin{center}
$\begin{array}{ccccccc}
\hline\hline
 \text{Eigenvalues} & 0 \ (2) & -\sqrt{x^2-2 x y+y^2+1} & \sqrt{x^2-2 x y+y^2+1} & -\sqrt{x^2+2 x y+y^2+1} & \sqrt{x^2+2 x y+y^2+1} \\
 \hline
 \mathrm{Zak \ phase} & \gamma_1+\gamma_2 & \gamma_3 & \gamma_4 &  \gamma_5 & \gamma_6
 \\
  & \pi  \left(2 -\frac{1}{(x+y)^2+1}-\frac{1}{(x-y)^2+1}\right) & \frac{\pi  \left(x^2-2 x y+y^2+2\right)}{2 \left(x^2-2 x
   y+y^2+1\right)}  & \frac{\pi  \left(x^2-2 x y+y^2+2\right)}{2 \left(x^2-2 x y+y^2+1\right)} & \frac{\pi  \left(x^2+2 x y+y^2+2\right)}{2
   \left(x^2+2 x y+y^2+1\right)} & \frac{\pi  \left(x^2+2 x y+y^2+2\right)}{2 \left(x^2+2 x y+y^2+1\right)} \\
   \hline\hline
\end{array}$\label{tab:3}
\end{center}
\end{table*}

\subsection{Non-trivial flat bands for uniform chains and Dexter-type topological phase transition}

\begin{figure*}[htbp]
\includegraphics[width=16cm, height=5.75cm, trim={0cm 0cm 0.0cm 0.0cm},clip]{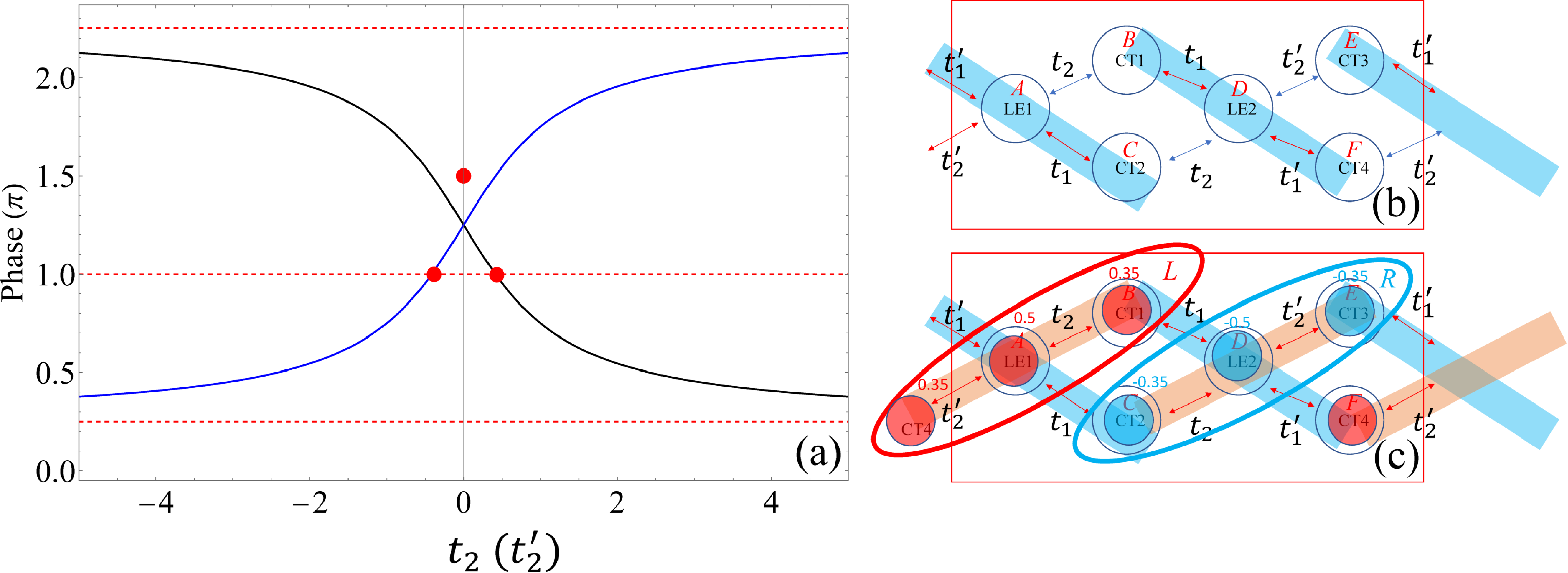}\\
\caption{(a) The Zak phases for the bottom two bands (the lowest band in blue) as a function of $t_2$ ($=t_2^\prime$) for a uniform chain ($t_1=t_1^\prime=1$). We have plotted the phases down to $-0.01$ ($0.01$) from the left-hand side (right-hand side). Notice that the phase could be $\pi$ even for a uniform chain at $t_2=t_2^\prime\simeq \pm(\sqrt{2}-1)$. The red points are the phases of $0$ and $\frac{3}{2}\pi$ for the case $t_1=t_1^\prime=1, t_2=t_2^\prime=0$. Therefore there is a phase transition at $x=0$, turning from $\frac{5}{2}\pi$ ($\frac{1}{2}\pi$) to $\frac{3}{2}\pi$ for the sum of the phases. On the other hand, when $x$ goes to infinity, the phases will approach $\frac{1}{4}\pi$ asymptotically. We show the coupling maps when $t_2=t_2^\prime = 0$ (b) and $t_2=t_2^\prime \neq 0$ (c), with $t_1=t_1^\prime=1$. We also show the coefficients for the eigenvector with the Zak phase of $\pi$ for $t_1=t_1^\prime=1, t_2=t_2^\prime\simeq\pm(\sqrt{2}-1)$. The coefficients have opposite signs in the red and blue circles, forming composite chiral sites.}\label{fig:3}
\end{figure*}

For the uniform-chain model, we have $t_1=t_1^\prime=1, t_2=t_2^\prime=x$. When $x \rightarrow \infty $, the phases for the bottom two bands will approach $\frac14\pi$ asymptotically, leading to a phase sum of $\frac{1}{2}\pi$, which is consistent with the uniform chain calculation shown in SI. This is different from the case for $t_1=t_1^\prime=0,t_2=t_2^\prime=1$ due to the symmetry and quantum interference. Two of the bands could be topologically nontrivial when $x\simeq\pm(\sqrt{2}-1)$, as shown in Fig.\ref{fig:3}(a). It is unexpected that we can have topologically nontrivial band even for a uniform chain in this model. For $x=\sqrt{2}-1$, we have checked the eigenvectors at the $\Gamma$-point, for which the coefficients for the states in the eigenvector with the Zak phase of $\pi$ is shown in Fig.\ref{fig:3}(a). The CCSs can be formed by the states in the red circle (left-hand) and in the blue circle (right-hand); their coefficients have opposite signs with same magnitude. In this case, the three states A, B, and F are exchanged simultaneously to D, E, and C, respectively, which would also respect the chiral symmetry. In addition, when $x$ approaching zero, there would be a two-fold degeneracy for the bottom two bands; the sum of their phases is $\frac{1}{2}\pi$ ($\frac{5}{2}\pi$), which is different from the two parameter model ($\frac{3}{2}\pi$), representing a phase transition due to symmetry breaking. As illustrated in Fig.\ref{fig:3}(b-c), the coupling map changes from stripes ($t_2=t_2^\prime=0$) to cross-nets (all the coupling turned on). Most importantly, we have also found the Zak phase of the zero-mode (or the middle for non-zero $d$) flat bands \textit{is always equal to $\pi$ for the uniform-chain with any energy gap $d$} according to our further calculations with $d\neq0$, whose Hamiltonian is shown in SI. This can also be interpreted as the inversion-symmetry protected topological phase.

\begin{figure*}[htbp]
\includegraphics[width=17cm, height=9.5cm, trim={0cm 0cm 0.0cm 0.0cm},clip]{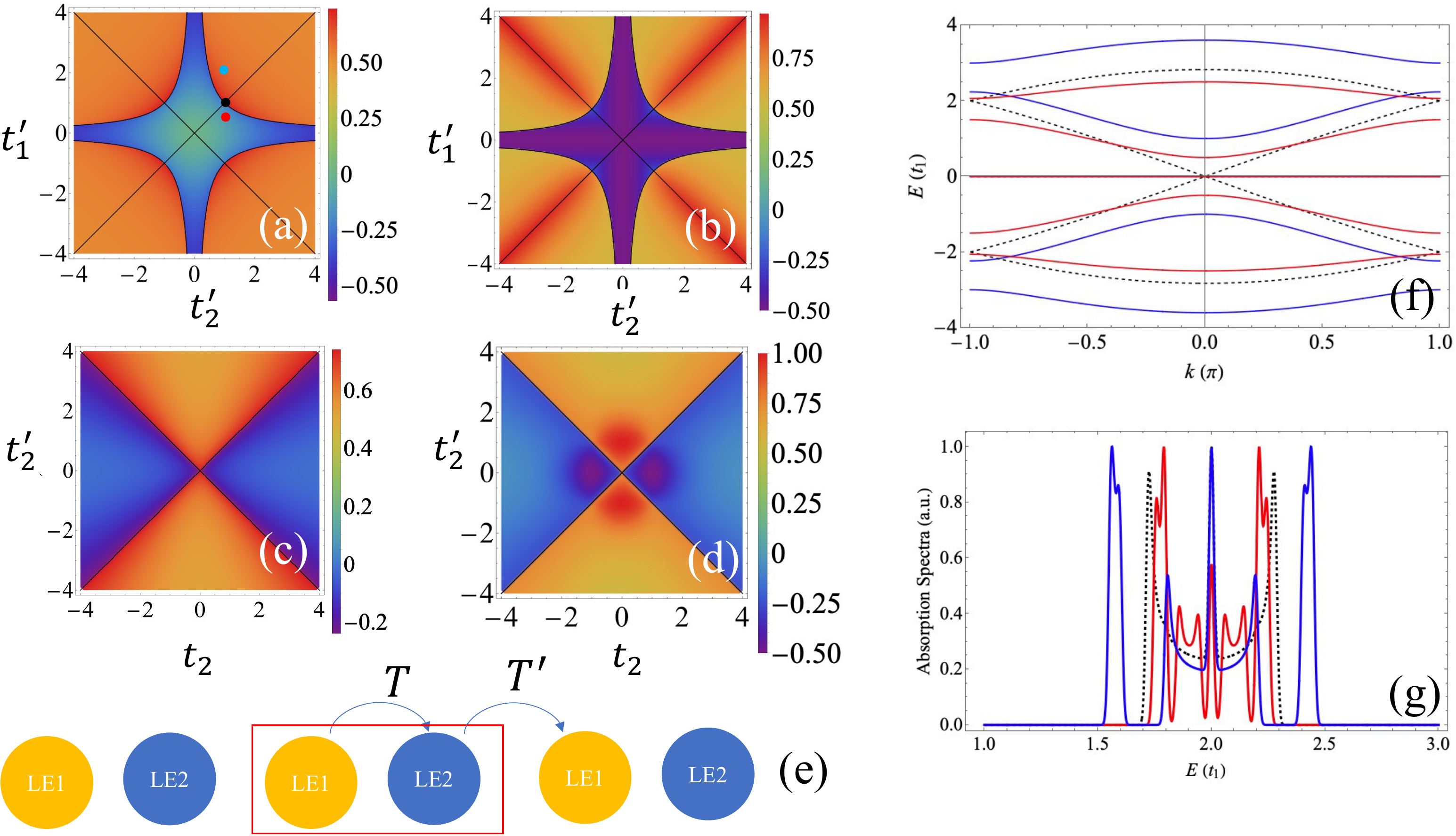}\\
\caption{The Zak phases for the bottom two bands computed for $t_1=t_2=1$ ($t_1=t_1^\prime=1$) are shown in (a-b) ((c-d)) as a function of $t_1^\prime$ and $t_2^\prime$ ($t_2$ and $t_2^\prime$). The phase transition mechanism is shown in (e). The band structures and corresponding absorption spectra are shown in (f) and (g) respectively, for the parameters at the blue ($t_1^\prime =1, t_2^\prime=2$), black dashed ($t_1^\prime =1, t_2^\prime=1$), and red ($t_1^\prime =1, t_2^\prime=\frac{1}{2}$) points indicated in (a).}\label{fig:t4_t12p}. 
\end{figure*}

We have also computed the case where $t_1 = t_2 = 1$ and $t_1^\prime = x, t_2^\prime = y$. The Zak phases for the bottom two bands as a function of $x$ and $y$ are shown in Fig.\ref{fig:t4_t12p}(a-b). At $\abs{t_1^\prime t_2^\prime} = 1$, there is a phase transition as suggested by the Zak phase going from $-\frac{\pi}{2}$ to $\frac{\pi}{2}$ induced by the topological order when tuning the hopping parameter continuously, which is more remarkable compared with the decoupling mechanism aforementioned. This topological phase transition is due to the Dexter-type excitonic hopping effect \cite{wu2008} between LE1 and LE2 where the effective hopping strengths are $T=t_1 t_2$ and $T^\prime=t_1^\prime t_2^\prime$ ($T$ is illustrated in Fig.\ref{fig:dexter}). Here the condition for the phase transition is that $\abs{T} = \abs{T^\prime}$, which is similar to the situation when the cross-cell hopping is equal to the intra-cell hopping in the SSH model, as shown in Fig.\ref{fig:t4_t12p}(e). This phase transition is also consistent with the previous work on the measurement of phase difference \cite{atala2013direct}. However, the states for the phase transition change to excitons, which is an entirely different from the conventional SSH model. This phase transition is further supported by our calculations for $t_1 = t_{1}^\prime = 1$ and $t_2 = x, t_2^\prime = y$ (Fig.\ref{fig:t4_t12p}(c-d)), showing a linear relationship between $t_2$ and $t_{2}^\prime$ at the phase transition. In Fig.\ref{fig:t4_t12p}(f-g), the band structures and corresponding normalized optical absorption spectra at the points ($t_1 = t_2=1, t_1^\prime =1, t_2^\prime=\frac{1}{2}$), ($t_1 = t_2 = 1,t_1^\prime = 1, t_2^\prime = 1$), and ($t_1 = t_2 = 1,t_1^\prime =1, t_2^\prime=2$) are shown to demonstrate the effect of phase transition. From the band structures, we can see the band gap opening and closing at $k=0$ and $k=\pm\pi$, which implies that these two phases are disconnected adiabatically. Our further calculations show that this type of phase transition can survive from the non-zero energy gap $d$ between LE and CT excitons, suggesting its robustness against perturbations. This type of phase transition has been reported previously \cite{atala2013direct} within the Rice-Mele model realized in the optical lattice. In relation to that, LE1 and LE2 are coupled by CT states, which would open a gap between them, corresponding to the energy offset $\Delta$ between the neighbouring sites in the Rice-Mele model. Notice that we also have another transition when $\abs{t_1^\prime} = \abs{t_2^\prime}$ owing to a change of the symmetry, which is fundamentally different. We have shown the results for this situation in Fig.2 in SI. The products of the hopping integrals have recently also been explored to identify the number of edge states for a four-band SSH model \cite{lee2022winding}.

Moreover, as the flat bands at energy zero are of great interest, we have also studied the phase sum for the scenario of $\{ x, y, 1, 1\}$, where generally $x\neq y$. Then the phase sum is equal to $\pi\abs{x^2-y^2}$, which is $\pi$ when $\abs{x^2-y^2}=1$. We have also derived the Zak phase formalism for a general scenario $\{ 1, x, y, z\}$ as shown in the SI, where in general ($x$,$y$, and $y$ none-zero and not equal to each other). Here we need to point out that the two-fold degeneracy for the flat bands is robust against any perturbation including non-zero $d$, which should be topological in the sense of topological order \cite{wen2016}.

\subsection{Edge states in finite-size chains}
\begin{figure*}[htbp]
\includegraphics[width=17.5cm, height=6.5cm, trim={0cm 0cm 0.0cm 0.0cm},clip]{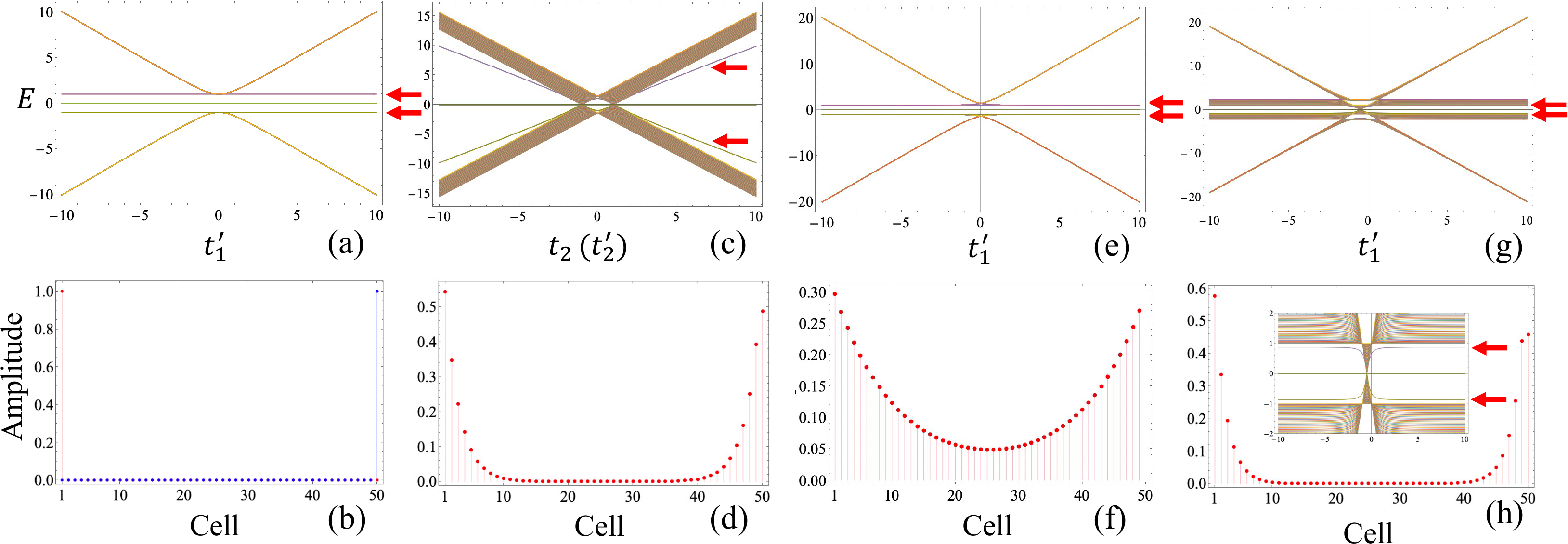}\\
\caption{Here we have computed the eigenvalues (a,c,e,g) and cell amplitudes (b,d,f,h) of a finite chain with 100 atomic sites. The red arrows point to the edge states chosen for the plots of amplitudes. The parameters for (a) is $\{1,0,t_1^\prime,0\}$, (b) is $\{1,0,2,0\}$, (c) is $\{1,t_2,1,t_2^\prime\}$, (d) is $\{1,2,1,2\}$, (e) is $\{0,1,t_1^\prime,\sqrt{1+t_1^{\prime 2}}\}$, (f) is $\{0,1,1,\sqrt{2}\}$, (g) is $\{1,1,t_1^\prime,t_1^\prime+1\}$, and (h) $\{1,1,2,3\}$. Notice that we have plotted the zoom-in spectra for (g) to illustrate the edge state is far off from the bulk states.} \label{fig:finitechain}. 
\end{figure*}

We have carried out calculations for the finite-site chains to illustrate the nature of the topological edge states related to the periodic-structure results.
The finite-chain calculations ($100$ sites) have been performed for four parameter sets discussed in the previous sections, i.e., $\{1,0,t_1^\prime,0\}$, $\{1,t_2,1,t_2^\prime\}$, $\{0,1,t_1^\prime,\sqrt{1+t_1^{\prime 2}}\}$, and $\{1,1,t_1^\prime,t_1^\prime+1\}$ in Fig.\ref{fig:finitechain} (a,c,e,g). Here we computed the eigenvalues as a function of the variable $x$. We have also computed the amplitude distributions among the cells for the edge states, as shown in Fig.\ref{fig:finitechain} (b,d,f,h). The zero-energy edge modes are only localized on one side of the chain, which is consistent with asymmetric nature of the periodic model. Apart from the zero-mode edge states, we have also found quasi-edge state with non-zero energy, for some of which the amplitude decays according to a power law. We have used red arrows to label the eigenvalues for these quasi-edge states in Fig.\ref{fig:finitechain}. For the cases $\{1,0,t_1^\prime,0\}$, $\{0,1,t_1^\prime,\sqrt{1+t_1^{\prime 2}}\}$ and $\{1,1,t_1^\prime,t_1^\prime+1\}$, the edge states are flat bands as pointed by the red arrows, whereas for $\{1,t_2,1,t_2^\prime\}$ (the uniform chain) the edge states appear off from the bulk states as shown in Fig.\ref{fig:finitechain}(c). The amplitude distributions for the edge states in the four parameter sets, computed for $\{1,0,2,0\}$, $\{1,2,1,2\}$, $\{0,1,1,\sqrt{2}\}$ and $\{1,1,2,3\}$, have shown a consistent trend. For the case $\{1,0,2,0\}$, the edge states are only concentrated on the edge cell; four edge states are associated (bonding/anti-bonding on each end). For the case $\{1,2,1,2\}$, the amplitudes decay exponentially from the edge towards the centre of the chain. For $\{1,1,2,3\}$, the amplitudes decay according to a power law.

\subsection{Potential experiments for the observation of excitonic edge states}
Our model has shown that there is strong indication that there could be edge states for any energy gap $d$ in the previous section. It could therefore be straightforward to observe the topological edge states in rather simple experimental setup. The only constraint is the isolation of states for the formation of excitons. The most convenient experimental materials platform is molecular chains in nanowires \cite{wang2010ultralong, zhao2023quantum}. Especially the highest occupied molecular orbitals (HOMO) and lowest unoccupied molecular orbitals (LUMO) could be well separated in porphyrin-like molecular chains, such as zinc-phthalocyanine (ZnPc) and copper phthalocyanine (CuPc) \cite{wang2010ultralong,guo2020new,liao2001electronic, marom2011electronic, wu2011theoretical}. Although the LUMO states are doubly degenerate ($e_{gx}$ and $e_{gy}$ under $D_{4h}$ symmetry), they could be decoupled approximately by symmetry. In addition, the HOMO ($a_{1u}$ state) is almost decoupled from the rest of orbitals. These great conditions will lead us to consider how we can observe the topological edge states in an obvious way. We could deposit or synthesis molecular nanowire such as ZnPc on Au(111) surfaces \cite{zhao2023quantum} then shed the light at the frequencies in the middle of the two peaks of the $Q$-bands \cite{wang2010ultralong,guo2020new}, then we could observe the light emission that should be on the one side of the chain ends. To observe the excitonic phase transition aforementioned, we might need to first realize the coupling map (Fig.\ref{fig:2}a) in optical lattice \cite{anisimovas2016}. We then need to perform the optical excitation according to the band structure shown in Fig.\ref{fig:t4_t12p} by using the methods outlined in Ref.\cite{atala2013direct}, in which varying the dimerisation type led to the observation of phase change of $\simeq \pi$. In optical lattice setup, we could tune the coupling product $t_1t_2$ to vary the dimerisation, thus observing the phase change due to the opposite winding directions of two types of dimerisations.  

\section{Conclusions}\label{sec:conclusions}
In summary, we have studied the topological properties of one-dimensional excitonic model that takes into account dimerization, the LE and CT excited states. We have found (i) a topological phase transition assisted by the Dexter electron exchange for the excitonic hopping and (ii) topologically nontrivial phase of $\pi$ could exist even for a uniform chain with any energy gap $d$, especially the robustness of the degeneracy for the zero-mode flat bands, implying topological order. In addition, we have also developed the concept of composite chiral sites for this type of systems when we have more than two states in a cell. Finite-chain calculations for 100 sites further confirm our periodic-structure calculations and illustrate the relevant edge states.

\section{Methods}\label{sec:methods}
We first solve the eigenvalues and eigenvectors ($u_n(k)$) for the Hamiltonian in the momentum space (eq.\ref{eq:cth}). Notice that when $d\neq 0$, the model becomes closer to the Rice-Mele model \cite{rice1982}. Then we compute the Zak phase by using the analytical and the numerical formalisms in the eq.\ref{eq:zp1} and eq.\ref{eq:zphase}, respectively. These have been used extensively in the previous work on the calculations of topological phases in one dimension \cite{zak1989berry, delplace2011zak, resta2000manifestations,le2020topological,zhu2020linear,zhao2023quantum}.

\begin{equation}\label{eq:zp1}
\gamma_n = i\int_{-\pi}^{\pi}dk\langle{u_n(k)}|{\partial_k}|u_n(k)\rangle.
\end{equation}
Here $u_n(k)$ is the Block wave function. We can still use eq.\ref{eq:zp1} to compute the Zak phase analytically if there are less than four non-zero parameters. However, for models with four non-zero parameters, we need to use the numerical formalism eq.\ref{eq:zphase} as below. Here by $n$-parameter, we mean there are $n$ non-zero parameters in the model.

\begin{equation}\label{eq:zphase}
    \gamma_n = \mathrm{Mod}(i \ln[\prod_{s=1}^M\scal{u_n(k_s)}{u_n(k_{s+1})}],2\pi),
\end{equation}
Here $u_n(k_s)$ is the eigvenvector of the Hamiltonian at $k_s$, $n$ labels the band, and $k_s$ run from $-\pi$ to $\pi$. We have tested the numerical robustness of our computational methods by using a series of different number of discritised points up to $1\times10^6$. We have found that $1\times10^5$ points are sufficient for the accuracy. 

We have also performed finite-chain calculations for 50 cells (in the last cell we have remove the CT3 and CT4 states such that the chain ends with LE states), equivalent to 100 atomic sites. We have computed the norm of the individual cells according to the eigenvectors.

When producing absorption spectra, we have followed the previous methods detailed in Ref.\cite{chen2021optoelectronic}. We assume the oscillator strengths for LE and CT to be 1 and 0.1, respectively. We have used a Gaussian-type broadening of $0.1 t_1$. As the model only takes into account the relative energy difference between LE and CT states, we have also include a rigid energy shift of $2 t_1$ in the calculations of absorption spectra.




\section*{Data Availability}
All the computer code and data that support the findings of this study are available from the corresponding author upon reasonable request.

\newpage
\bibliography{cthbib}

\section*{Acknowledgements}
JHZ and JC acknowledge the funding from the National Natural Science Foundation of China under Grant No. 92165101. We thank the inspiring discussions with colleagues from UCL and China. WW wishes to acknowledge the support of the UK Research Councils under Programme Grant EP/M009564/1, the EU Horizon 2020 Project Marketplace (No. 760173), and UK Science and Technology Facilities Council for funding.

\section{Author Contributions}

All the authors contributed to the concept of the paper. WW and JHZ performed the theoretical analysis. All the authors wrote the paper. 

\section*{Competing Interests}
The authors declare no competing interests.

\end{document}